\documentclass{iopconfser}
\usepackage[utf8]{inputenc}
\usepackage{array}
\usepackage{graphicx}
\usepackage{amsmath}
\usepackage{amsfonts}
\usepackage{amssymb}
\usepackage{caption}
\usepackage{subcaption}
\usepackage{float}
\usepackage{cite}
\usepackage{fullpage}
\usepackage{enumerate}
\usepackage{setspace}
\usepackage{float}
\usepackage{cite}
\usepackage[colorlinks=true, allcolors=blue]{hyperref}
\usepackage{ragged2e}
\begin{document}
\pagestyle{empty}
\title{Fokas-Lenells Derivative nonlinear Schr\"odinger equation its associated soliton surfaces and Gaussian curvature }

\author{Sagardeep Talukdar$^{1*}$, Riki Dutta$^{1}$, Gautam Kumar Saharia$^{1}$ and Sudipta Nandy$^{1}$}

\affil{$^1$Department of Physics, Cotton University, Guwahati-781001, Assam, India}

\email{phy2091005\_sagardeep@cottonuniversity.ac.in}

\begin{abstract}
\justifying	
One of the most important tasks in mathematics and physics is to connect differential geometry and nonlinear differential equations. In the study of nonlinear optics, integrable nonlinear differential equations such as the nonlinear Schr\"odinger equation (NLSE) and higher-order NLSE (HNLSE) play crucial roles. Because of the medium's balance between dispersion and nonlinearity, all of these systems display soliton solutions. The soliton surfaces, or manifolds, connected to these integrable systems hold significance in numerous areas of mathematics and physics. We examine the use of soliton theory in differential geometry in this paper. We build the two-dimensional soliton surface in the three-dimensional Euclidean space by taking into account the Fokas-Lenells Derivative nonlinear Schrödinger equation (also known as the gauged Fokas-Lenells equation). The same is constructed by us using the Sym-Tafel formula. The first and second fundamental forms, surface area, and Gaussian curvature are obtained using a lax representation of the gauged FLE.
\end{abstract}
\vspace{10mm}
\section{Introduction}
\justifying
The propagation of ultrashort pulses in nonlinear media is described by the Fokas-Lenells equation (FLE) \cite{Fokas, Lenells, FL}, one of the four categories of integrable equations. The nonlinear Schr\"odinger equation (NLSE)\cite{Hasegawa, Serkin00,Nandy}, derivative NLSE (DNLSE) \cite{Kaup}, HNLSE \cite{Hirota, SSE} are the other three. The study of localized waves in nonlinear optical media involves the application of all four equations. \\
\indent A feature shared by all these physical systems is the appearance of solitons, which result from a dispersive and nonlinear term balance in the wave equations \cite{Agrawal}. The inclusion of a spatio-temporal dispersion term in addition to the group velocity dispersion term accounts for the significance of FLE when compared to NLSE. It appears when input ultrashort pulse propagation in optical fibers is taken into account along with higher order effects like self-steepening, Raman scattering, and third order dispersion.\\
\vspace{2mm}
\indent The dimensionless form of FLE \cite{Fokas, Lenells} :
\begin{align} 
	\label{FWDU}
	iU_t  + \gamma_1 U_{xx} + \gamma_3 |U|^2 U + \gamma_2 ( i \gamma_3  |U|^2U_x-U_{xt}) = 0 
\end{align}
in which $U$ represents an optical field's envelope function. The group velocity dispersion term and the nonlinear term are the second and third terms of eq. \ref{FWDU}. The last term in eq. \ref{FWDU} is the spatio-temporal dispersion term, and the second last term is the intensity dependent contribution to the group velocity, which in turn causes the phenomenon of self-steepening\cite{Agrawal,BOYD2003533}. It can explain how an optical field propagates in an ultrashort pulse when it is allowed to diffract along either of the two directions—that is, longitudinally or transversely.  An important fact about the FLE is that a gauge transformation of eq. \ref{FWDU} belongs to the  hierarchy   of integrable DNLSE  \cite{Matsuno1,Matsuno2,talukdar,dutta2023fokas,saharia, sagar2, dutta2} which we refer to as gauged FLE in this paper. The eq. \ref{FWDU} then gets transformed into the form,
\begin{align} 
	\label{FLE}
	q_{\xi \tau} = q - 2i |q|^2 q_\xi      
\end{align}
\indent In recent decades, significant strides have been made in the field of differential geometry pertaining to surfaces within Euclidean space. The profound interrelation between classical differential geometry and soliton theory has been firmly established subsequent to the inception of the Bäcklund-Darboux transformation\cite{rogers}. \\
\indent The idea of associating a class of surfaces or manifolds with a given integrable system of nonlinear equations through the Sym-Tafel formula is well known\cite{sym}. It allows us to explicitly construct of surfaces from the fundamental forms (FFs). The FFs play an important role in understanding the geometric aspects of the surface in analogue black hole study. While research have been done to investigate associated surfaces for different integrable systems\cite{myrzakulov1,myrzakulov2}, the creation of associated soliton surfaces for the Fokas-Lenells Derivative nonlinear Schrödinger equation or gauged FLE remains, to the best of our knowledge yet not carried out. In this manuscript, our focus is on the construction of soliton surfaces linked to the gauged FLE utilizing the Sym-Tafel methodology. Our investigation starts from the LAX pair of integrable systems to the study of surface geometry within Euclidean spaces. 
\\
\indent	The structure of the manuscript is as follows. In the first section we introduce the gauged FLE and the associated LAX pair of it obtained in \cite{talukdar}. In the second section, we obtain the set of fundamental forms for it. In the third section, we investigate the surface area, in the fourth section we calculate the Gaussian curvature. The fifth section will be the concluding one.
Since, eq. \ref{FLE} is integrable in Liouville sense, it admits LAX pair.  The linear transformation equation for gauged FLE in terms of  Lax pair is given by,

\begin{align}
	\label{LaxFL}
	\partial_\xi {\Psi} = L { \Psi} \\
	\partial_\tau { \Psi} = M { \Psi}
\end{align}

where $\bf{\Psi}$ is a two component vector field and is the function of $\xi$, $\tau$ and spectral parameter $\zeta$ expressed as,
\begin{align}
	\label{Psi}
	{ \Psi}= ({\Psi_1} \ {\Psi_2})^T
\end{align}

and $L$ and $M$ are given by
\begin{align}
	\label{Lax2}
	L = \frac{-i\zeta^2 }{2} \ \sigma   + \zeta \  \partial_\xi u           \\
	M= \frac{i }{2  } \zeta^{-2} \ \sigma -  {i}{\zeta^{-1}}\ \sigma u + i\sigma u^2
\end{align} 

$\sigma$  and $u$ are $2\times 2 $ matrices, defined as follows

$ \quad \sigma = \left(  \begin{tabular}{c c}
	1 & 0 \\ 
	0 & -1  \\ 
\end{tabular} \right) , \quad \quad
u = \left(  \begin{tabular}{c c}
	0 & $q$ \\ 
	$-q^*$ & 0   \\
\end{tabular} \right) $ \\

\section{Fundamental Form}
\vspace{2mm}
In this section, we introduce the Sym-Tafel formula and find the first and second fundamental form. The Sym Tafel formula behaves as a string connecting the theory of solitons and classical geometry as it relates the two-dimensional surfaces and integrable equations as given by,
\begin{align}\label{eq:STF}
	\textbf{r}=\psi^{-1}\psi_\zeta
\end{align}
where $\textbf{r}=\textbf{r}(\xi,\tau)$ denotes the position vector of a point $P$ on a surface $\Sigma$ in $\mathbb{R}^3$.
\subsection{The first fundamental form (1FF) of the surface}
\vspace{1mm}
The 1FF or the metric tensor of $\Sigma$ primarily measures infinitely small arcs on a surface. 1FF of the smooth surface P is obtained by the scalar product of the total differential of position vector \textbf{r} i.e.,
\begin{align}
	I=d\textbf{r}.d\textbf{r}
\end{align}
where $\textbf{dr} = \textbf{r}_\xi d\xi + \textbf{r}_\tau d\tau$.
\begin{align}
	I=dr^2&={\textbf{r}^2}_\xi d\xi^2+2 \textbf{r}_\xi \textbf{r}_\tau d\xi d\tau+{\textbf{r}^2}_\tau d\tau^2\\
	or I&=E  d\xi^2+ 2 F d\xi d\tau+ G d\tau^2 \label{eq:1ff}
\end{align}
where $H^2=EG-F^2$ is the metric discriminant of the 1FF satisfying the condition $H>0$. The coefficients are given by,
\begin{align*} 
	E&={\textbf{r}^2}_\xi \\
	F&= \textbf{r}_\xi \textbf{r}_\tau\\
	G&={\textbf{r}^2}_\tau
\end{align*}
Now, using eq. \ref{eq:STF} we calculate the above coefficients as,
\begin{align}
	 \textbf{r}_\xi&= (\psi^{-1}\psi_\zeta)_\xi=\psi^{-1}L_\zeta \psi \\
	  \textbf{r}_\tau&= (\psi^{-1}\psi_\zeta)_\tau=\psi^{-1}M_\zeta \psi
	 \end{align}
 Therefore, the coefficients take the form as,
 \begin{align*}
 	E&= {\textbf{r}_\xi}^2=(\psi^{-1}L_\zeta \psi)(\psi^{-1}L_\zeta \psi)=-\frac{1}{2} Tr({L}_\zeta^2)\\
 	F&= \textbf{r}_\xi \textbf{r}_\tau=(\psi^{-1}L_\zeta \psi)(\psi^{-1}M_\zeta \psi)=-\frac{1}{2} Tr({L}_\zeta M_\zeta)\\
 	G&= {\textbf{r}_\tau}^2=(\psi^{-1}M_\zeta \psi)(\psi^{-1}M_\zeta \psi)=-\frac{1}{2} Tr({M}_\zeta^2)\\
 \end{align*}
In terms of matrix $L$ and $M$ eq. \ref{eq:1ff} takes the form as,
\begin{align}
	I=-\frac{1}{2}[ Tr({L}_\zeta^2){d\zeta}^2+2 Tr({L}_\zeta M_\zeta)d\xi d\tau+Tr({M}_\zeta^2) d\tau^2]
\end{align}
Now, we differentiate the matrix operators $L$ and $M$ with respect to the spectral parameter $\zeta$. We find,
\begin{align*}
	L_\zeta=\begin{pmatrix} -i\zeta & q_\xi \\
		-q_\xi^* & i\zeta  
	\end{pmatrix}; 
\quad \quad
	L_\zeta^2=\begin{pmatrix} -\zeta^2-|q_\xi|^2 & 0 \\
		0 & -\zeta^2-|q_\xi|^2  
	\end{pmatrix}\\
M_\zeta=\begin{pmatrix} \frac{-i}{\zeta^3} & \frac{-iq}{\zeta^2} \\
	\frac{-iq^*}{\zeta^2} & \frac{i}{\zeta^3}  
\end{pmatrix}; 
\quad \quad
M_\zeta^2=\begin{pmatrix} \frac{-1}{\zeta^6}-\frac{|q|^2}{\zeta^4} & 0 \\
	0 & \frac{-1}{\zeta^6}-\frac{|q|^2}{\zeta^4} 
\end{pmatrix}
\end{align*}
\begin{align*}
 M_\zeta N_\zeta=\begin{pmatrix} \frac{-1}{\zeta^2}-\frac{i q_\xi q^*}{\zeta^2} & \frac{-q}{\zeta}+\frac{i q_\xi}{\zeta^3} \\
	\frac{q^*}{\zeta}+\frac{i q_\xi^*}{\zeta^3} & \frac{-1}{\zeta^2}+\frac{i q_\xi^* q}{\zeta^2}
\end{pmatrix}\\
\end{align*}
Therefore, the values of the coefficients are,
 \begin{align}\label{eq:E}
	E&= -\frac{1}{2} Tr({L}_\zeta^2)=(\zeta^2+|q_\xi|^2)\\ \label{eq:F}
	F&= -\frac{1}{2} Tr({L}_\zeta M_\zeta)=\frac{1}{\zeta^2}+\frac{i}{\zeta^2}(q_\xi q^*-q q_\xi^*)\\ \label{eq:G}
	G&= -\frac{1}{2} Tr({M}_\zeta^2)=(\frac{1}{\zeta^6}+\frac{|q|^2}{\zeta^5})
\end{align}
Substituting the above values in eq. \ref{eq:1ff} we get,
\begin{align}
	I=(\zeta^2+|q_\xi|^2)d\xi^2+2[\frac{1}{\zeta^2}+\frac{i}{\zeta^2}(q_\xi q^*-q q_\xi^*)]d\xi d\tau+(\frac{1}{\zeta^6}+\frac{|q|^2}{\zeta^5})d\tau^2
\end{align}

Thus, the above equation represents the 1FF or the metric tensor.

\subsection{Second fundamental form (2FF) of the surface}
\vspace{2mm}
The 2FF describes how the surface deviates from the tangent plane and completely determines the curvature of the surface. Let, \textbf{$r_\xi$} and \textbf{$r_\tau$} be the tangent vectors to $\Sigma$ at $P$, then the unit normal to $\Sigma$ is denoted as $\hat{n}$. The 2FF then is represented as the scalar product of the vector $dr$ and $\hat{n}$. 
Mathematically,
\begin{align}\label{eq:2ff}
	II=-d^2\textbf{r}.\hat{n}
\end{align}
where,
 \begin{align}
 \hat{n}=\frac{\textbf{r}_\xi\times \textbf{r}_\tau}{|\textbf{r}_\xi\times \textbf{r}_\tau|}
\end{align}
and $d^2 \textbf{r}=\textbf{r}_{\xi \xi} d\xi^2+2 \textbf{r}_\xi \textbf{r}_\tau d\xi d\tau+\textbf{r}_{\tau\tau} d\tau^2$. 
So, eq. \ref{eq:2ff} takes the form as,
\begin{align}
	II=\textbf{r}_{\xi \xi}.\hat{n} d\xi^2+2{\textbf{r}_\xi \textbf{r}_\tau}.\hat{n} d\xi d\tau+\textbf{r}_{\tau\tau}.\hat{n} d\tau^2\\ \label{eq:2ff2}
	or II=E'd\zeta^2+2F' d\zeta d\tau+G'd\tau^2
\end{align}
where,
\begin{align}\label{eq:E'}
	E'&=\textbf{r}_{\xi \xi}.\hat{n}=\frac{1}{2}Tr(\textbf{r}_{\xi \xi}\hat{n})\\\label{eq:F'}
	F'&={\textbf{r}_\xi \textbf{r}_\tau}.\hat{n}=\frac{1}{2}Tr({\textbf{r}_{\xi\tau}}\hat{n})\\
	\label{eq:G'}
	G'&=\textbf{r}_{\tau\tau}.\hat{n}=\frac{1}{2}Tr(\textbf{r}_{\tau \tau}\hat{n})\\
	\hat{n}&=\frac{\psi^{-1}[L_\zeta,M_\zeta]\psi}{\sqrt{\frac{1}{2} Tr([L_\zeta,M_\zeta]^2)}}
	\end{align}
Again, the second derivative of eq. \ref{eq:STF} can be calculated to be,
\begin{align}\label{eq:sd}
	\textbf{r}_{\xi \xi}&= \psi^{-1}(L_{\zeta\xi}+[L_\zeta,L])\psi\\
	\textbf{r}_{\tau \tau}&= \psi^{-1}(M_{\zeta\tau}+[M_\zeta,M])\psi\\
	\textbf{r}_{\xi \tau}&= \psi^{-1}(L_{\zeta\tau}+[L_\zeta,M])\psi\\
	\end{align}
Substituting eq. \ref{eq:sd} in eq. \ref{eq:E'}-\ref{eq:G'} the coefficients take the form as,
\begin{align}\label{eq:E''}
		E'&=\frac{1}{2}\frac{Tr((L_{\zeta\xi}+[L_\zeta,L])[L_\zeta,M_\zeta])}{\sqrt{\frac{1}{2} Tr([L_\zeta,M_\zeta]^2)}}\\ \label{eq:F''}
	F'&=\frac{1}{2}\frac{Tr((L_{\zeta\tau}+[L_\zeta,M]([L_\zeta,M_\zeta])}{\sqrt{\frac{1}{2} Tr([L_\zeta,M_\zeta]^2)}}\\ 
	     \label{eq:G''}
	G'&=\frac{1}{2}\frac{Tr((M_{\zeta\tau}+[M_\zeta,M])[L_\zeta,M_\zeta])}{\sqrt{\frac{1}{2} Tr([L_\zeta,M_\zeta]^2)}}
\end{align}
Now, we try to find the commutator relations to calculate the 2FF,
\begin{align}\label{eq:coef1}
[L_\zeta,L]= \left(
\begin{array}{cc}
	0 & -i \zeta ^2 q_{\xi } \\
	-i \zeta ^2 q^*_{\xi } & 0 \\
\end{array}
\right)
\end{align}
\begin{align}\label{eq:coef2}
 [l_\zeta,M]=\left(
 \begin{array}{cc}
 	\frac{i q q^*_{\xi }}{\zeta }+\frac{i q_{\xi } q^*}{\zeta } & 2 q+q_{\xi } \left(-\frac{i}{2 \zeta ^2}+i q q^*\right)-q_{\xi } \left(\frac{i}{2 \zeta ^2}-i q q^*\right) \\
 	-2 q^*+q^*_{\xi } \left(-\frac{i}{2 \zeta ^2}+i q q^*\right)-q^*_{\xi } \left(\frac{i}{2 \zeta ^2}-i q q^*\right) & -\frac{i q q^*_{\xi }}{\zeta }-\frac{i q_{\xi } q^*}{\zeta } \\
 \end{array}
 \right)
\end{align}
\begin{align}\label{eq:coef3}
[M_\zeta,M]=\left(
	\begin{array}{cc}
		0 & \frac{i q \left(\frac{i}{2 \zeta ^2}-i q q^*\right)}{\zeta ^2}-\frac{i q \left(-\frac{i}{2 \zeta ^2}+i q q^*\right)}{\zeta ^2} \\
		\frac{i q^* \left(-\frac{i}{2 \zeta ^2}+i q q^*\right)}{\zeta ^2}-\frac{i q^* \left(\frac{i}{2 \zeta ^2}-i q q^*\right)}{\zeta ^2} & 0 \\
	\end{array}
	\right)
\end{align}
\begin{align}\label{eq:coef4}
[L_\zeta,M_\zeta]=\left(
\begin{array}{cc}
	-\frac{i q q^*{}_{\xi }}{\zeta ^2}-\frac{i q_{\xi } q^*}{\zeta ^2} & -\frac{2 q}{\zeta } \\
	\frac{2 q^*}{\zeta } & \frac{i q q^*_{\xi }}{\zeta ^2}+\frac{i q_{\xi } q^*}{\zeta ^2} \\
\end{array}
\right)
\end{align}
Again,
\begin{align}\label{eq:coef5}
	L_{\zeta  \xi }=\left(
	\begin{array}{cc}
		0 & q_{\xi \xi } \\
		-q^*_{\xi \xi } & 0 \\
	\end{array}
	\right)
	\quad \quad 
	L_{\zeta  \tau }=\left(
	\begin{array}{cc}
		0 & q_{\xi \tau } \\
		-q^*_{\xi \tau } & 0 \\
	\end{array}
	\right) \quad \quad
	M_{\zeta \tau }=\left(
	\begin{array}{cc}
		0 & -\frac{i q_{\tau }}{\zeta ^2} \\
		-\frac{i q^*_{\tau }}{\zeta ^2} & 0 \\
	\end{array}
	\right)
\end{align}
Substituting the values obtained in eqs. \ref{eq:coef1}-\ref{eq:coef5} into eqs. \ref{eq:E''}-\ref{eq:G''} we obtain the coefficients of the 2FF as,
\begin{align}
	E'=\frac{\sqrt{2} \left(\zeta ^2 q^* \left(q_{\xi \xi }-i \zeta ^2 q_{\xi }\right)+q^*_{\xi \xi } \left(\zeta ^2 q-i q_{\xi }\right)+q^*_{\xi } \left(2 \zeta ^2 q_{\xi }+i \left(\zeta ^4 q+q_{\xi \xi }\right)\right)\right)}{\zeta ^3 \sqrt{\left(\frac{i q q^*_{\xi }}{\zeta ^2}+\frac{i q^* q_{\xi }}{\zeta ^2}\right)^2+\left(-\frac{i q q^*_{\xi }}{\zeta ^2}-\frac{i q^* q_{\xi }}{\zeta ^2}\right)^2}} 
\end{align}
\begin{align}
	F'=\frac{\sqrt{2} \left(\zeta ^2 q^2 \left(q^*\right)_{\xi }^2+q^*_{\xi } \left(\left(2-2 \zeta ^2 q q^*\right) q_{\xi }+i \zeta ^2 \left(q \left(3-2 \zeta ^2 q q^*\right)+q_{\xi \tau }\right)\right)\right)}{\zeta ^5 \sqrt{\left(\frac{i q q^*_{\xi }}{\zeta ^2}+\frac{i q^* q_{\xi }}{\zeta ^2}\right)^2+\left(-\frac{i q q^*_{\xi }}{\zeta ^2}-\frac{i q^* q_{\xi }}{\zeta ^2}\right)^2}} \nonumber \\
	+\frac{\sqrt{2} \left(\zeta ^2 \left(q^*{}_{\xi \tau } \left(\zeta ^2 q-i q_{\xi }\right)+q^* \left(\zeta ^2 \left(q_{\xi \tau }+4 q\right)+q_{\xi } \left(q^* \left(q_{\xi }+2 i \zeta ^2 q\right)-3 i\right)\right)\right)\right)}{\zeta ^5 \sqrt{\left(\frac{i q q^*_{\xi }}{\zeta ^2}+\frac{i q^* q_{\xi }}{\zeta ^2}\right)^2+\left(-\frac{i q q^*{}_{\xi }}{\zeta ^2}-\frac{i q^* q_{\xi }}{\zeta ^2}\right)^2}} 
\end{align}
\begin{align}
	G'=\frac{\sqrt{2} \left(\zeta ^2 q^*{}_{\tau } \left(q_{\xi }+i \zeta ^2 q\right)+q^*_{\xi } \left(\zeta ^2 q_{\tau }+i q \left(2 \zeta ^2 q q^*+1\right)\right)+q^* \left(-i \left(2 \zeta ^2 q q^*+1\right) q_{\xi }+\zeta ^4 \left(4 q^2 q^*-i q_{\tau }\right)+2 \zeta ^2 q\right)\right)}{\zeta ^7 \sqrt{\left(\frac{i q q^*_{\xi }}{\zeta ^2}+\frac{i q^* q_{\xi }}{\zeta ^2}\right)^2+\left(-\frac{i q q^*{}_{\xi }}{\zeta ^2}-\frac{i q^* q_{\xi }}{\zeta ^2}\right)^2}}
\end{align}
Substituting the above values in eq. \ref{eq:2ff2} we obtain the 2FF.
\section{Surface Area}
"If a surface within is defined by a parametrically smooth function \( r(\xi, \tau) \), where the parameters $\xi $ and $\tau$ vary within the plane, then the surface area $A$ can be represented through the double integral,
\begin{align}
A&=\int \int {H} d\xi d\tau \\
\implies A&=\int \int |r_\xi \times r_\tau|d\xi d\tau \\ \label{eq:A}
\implies A&=\int \int \sqrt{EG-F^2}d\xi d\tau
\end{align}
where $E$, $G$ and $F$ are given by the relations eqs. \ref{eq:E}-\ref{eq:G}. \\
Now, substituting the values of $E$, $F$ and $G$ into eq. \ref{eq:A} we get,
\begin{align} \nonumber
 A&=\int \int \sqrt{\left(\frac{1}{\zeta ^6}+\frac{q q^*}{\zeta ^5}\right) \left(\zeta ^2+q_{\xi } q^*_{\xi }\right)-\left(\frac{1}{\zeta ^2}+\frac{i \left(q^* q_{\xi }-q q^*_{\xi }\right)}{\zeta ^2}\right)^2}d\xi d\tau\\ 
	\nonumber
	\implies A&=\int \int \sqrt{\frac{\left(\zeta  |q|^2 +1\right) \left(\zeta ^2+|q_{\xi }|^2 \right)+\zeta ^2 \left(q q^*_{\xi }-q^* q_{\xi }+i\right)^2}{\zeta ^6}}d\xi d\tau
\end{align}
The surface area takes the form as,
\begin{align}
		 A&=\frac{1}{\zeta^3} \int \int \sqrt{\left(\zeta |q|^2 +1\right) \left(\zeta ^2+|q_{\xi }|^2 \right)+\zeta ^2 \left(q q^*_{\xi }-q^* q_{\xi }+i\right)^2} d\xi d\tau
\end{align}

\section{Gaussian Curvature}
The Gaussian curvature is defined to be the product of the two principal curvatures $\kappa_1$ and $\kappa_2$ or the ratio of the discriminant of 1FF to the 2FF. Now, let us introduce two symmetric $2\times2$ matrices from eq. \ref{eq:1ff} and \ref{eq:2ff2} respectively.
\begin{align}
	g_1=\begin{pmatrix} E & F \\
		F & G  
	\end{pmatrix} 
\quad \quad
	g_2=\begin{pmatrix} E' & F' \\
	F' & G'  
\end{pmatrix} 
\end{align} 
So, the principal curvatures are the eigenvalues of $g_1^{-1}g_2$ \cite{pressley}. Therefore, the determinant of the matrix $g_1^{-1}g_2$ is the product of the two principal curvatures i.e. $\kappa_1 \kappa_2$. \\
Mathematically, the Gaussian curvature becomes,
\begin{align*} 
	 K&=det(g_1^{-1}g_2)\\ 
	 \implies K&=det(g_1^{-1})det(g_2)\\ 
	 \implies K&=\frac{E'G'-F'^2}{EG-F^2}
\end{align*} 
Substituting the values of the 1FF and 2FF coefficients, the Gaussian curvature can be calculated. 

\section{Conclusion}
 In this manuscript, the Sym-Tafel formula has been utilized to construct the 1FF and 2FF for the gauged FLE. The associated coefficients for the same were computed. On the smooth surface, the 1FF or metric provides the small arc length between two points, and the 2FF provides a quadratic form on the tangent plane. It displays how the surface deviates from its tangent plane. When analyzing the geometric characteristics of a surface, both the 1FF and 2FF are crucial. Utilizing the LAX pair for gauged FLE, which was previously constructed in \cite{talukdar}, the Sym-Tafel formula derives the fundamental forms. Subsequently, we calculated the surface area and used the quadratic coefficients of the 1FF and 2FF to calculate the Gaussian curvature. The current study, in our opinion, will be beneficial for the study of analog black holes as well as several other areas of physics and mathematics.
 
\section{Acknowledgement} 
S. Talukdar and R. Dutta acknowledge Department
of Science \& Technology, Govt. of India for Inspire fellowship,
grant nos. DST/INSPIRE Fellowship/2020/IF200278 and DST/
INSPIRE Fellowship/2020/IF200303.


\end{document}